\documentclass[sigconf]{acmart}

\usepackage{graphicx}   % for including images
\usepackage{subcaption} % for subfigures
\usepackage{soul}
\usepackage{footmisc}
\usepackage{flushend}
\usepackage{xfrac}

\AtBeginDocument{%
  }

\copyrightyear{2023} 
\acmYear{2023} 
\setcopyright{acmlicensed}\acmConference[HotCarbon '23]{2nd Workshop on Sustainable Computer Systems}{July 9, 2023}{Boston, MA, USA}
\acmBooktitle{2nd Workshop on Sustainable Computer Systems (HotCarbon '23), July 9, 2023, Boston, MA, USA}
\acmPrice{15.00}
\acmDOI{10.1145/3604930.3605709}
\acmISBN{979-8-4007-0242-6/23/07}

\begin{document}

\title{The War of the Efficiencies: Understanding the Tension between Carbon and Energy Optimization}

\author{Walid A. Hanafy, Roozbeh Bostandoost, Noman Bashir, David Irwin, Mohammad Hajiesmaili, and Prashant Shenoy}
\affiliation{%
University of Massachusetts Amherst
\country{USA}
}

\renewcommand{\shortauthors}{Hanafy, et al.}

\begin{abstract}
Major innovations in computing have been driven by scaling up computing infrastructure, while aggressively optimizing operating costs. 
The result is a network of worldwide datacenters that consume a large amount of energy, mostly in an energy-efficient manner. 
Since the electric grid powering these datacenters provided a simple and opaque abstraction of an unlimited and reliable power supply, the computing industry remained largely oblivious to the carbon intensity of the electricity it uses. 
Much like the rest of the society, it generally treated the carbon intensity of the electricity as constant, which was mostly true for a fossil fuel-driven grid. 
As a result, the cost-driven objective of increasing energy-efficiency --- by doing more work per unit of energy --- has generally been viewed as the most carbon-efficient approach. 
However, as the electric grid is increasingly powered by clean energy and is exposing its time-varying carbon intensity, the most energy-efficient operation is no longer necessarily the most carbon-efficient operation.
There has been a recent focus on exploiting the flexibility of computing's workloads---along temporal, spatial, and resource dimensions---to reduce carbon emissions, which comes at the cost of either performance or energy efficiency.
In this paper, we discuss the trade-offs between energy efficiency and carbon efficiency in exploiting computing's flexibility and show that blindly optimizing for energy efficiency is not always the right approach.
\end{abstract}
\keywords{Carbon Efficiency, Energy Efficiency, Sustainable Computing}

\begin{CCSXML}
<ccs2012>
   <concept>
       <concept_id>10010520.10010521.10010537.10003100</concept_id>
       <concept_desc>Computer systems organization~Cloud computing</concept_desc>
       <concept_significance>500</concept_significance>
       </concept>
   <concept>
       <concept_id>10010583.10010662.10010663.10010666</concept_id>
       <concept_desc>Hardware~Renewable energy</concept_desc>
       <concept_significance>500</concept_significance>
       </concept>
   <concept>
       <concept_id>10003456.10003457.10003458.10010921</concept_id>
       <concept_desc>Social and professional topics~Sustainability</concept_desc>
       <concept_significance>500</concept_significance>
       </concept>
 </ccs2012>
\end{CCSXML}

\ccsdesc[500]{Computer systems organization~Cloud computing}
\ccsdesc[500]{Hardware~Renewable energy}
\ccsdesc[500]{Social and professional topics~Sustainability}

\settopmatter{printfolios=true}
\maketitle

\section{Introduction}
\label{sec:introduction}
The demand for computing has experienced rapid growth and is expected to accelerate even further. 
However, the increase in computing demand has not resulted in a proportional increase in energy demand so far~\cite{iea-demand}.
The growth in computing's energy consumption has been kept in check by massive gains in algorithmic efficiency, measured in cycles per unit of work, of its software and energy efficiency, measured in energy consumption per cycle, of its hardware~\cite{hotcarbon-hotair}. 
However, as the algorithmic and energy efficiency gains slow down, an increase in computing demand directly increases the energy demand. 
A conservative estimate projects that the energy consumption of datacenters will increase by at least 10\% per year till 2030~\cite{datacenter-demand}, much higher than an estimated increase of 1.65\% per year in 2010s~\cite{masanet}.
As society has begun to recognize the environmental impact of our activities, reducing the carbon footprint of this accelerating energy demand has attracted significant attention from academic researchers~\cite{ecovisor, lechowicz2023online, wait-awhile, switzer2023junkyard} and industry leaders~\cite{acun2023carbon, radovanovic2021carbonaware}.

%noman's draft
The carbon footprint of computing depends on the computing's carbon efficiency, denoted as $\eta_\texttt{C}$, which is calculated by dividing the computing's energy efficiency $\eta_\texttt{E}$, measured as work done per unit of energy, by the energy's carbon intensity $c$, measured as the amount of emitted greenhouse gases (GHG) per kWh of energy.
Traditionally, the electric grid has been powered by fossil fuels such as coal,  and oil, which have  similar carbon intensities of 1038~$g.CO_2eq/kWh$ and 1106~$g.CO_2eq/kWh$~\cite{carbon-emissions-by-type}.
Furthermore, even if energy's carbon intensity slightly varied across space and time, it was invisible to electricity consumers due to the simple and opaque abstraction exposed by the grid. 
As a result, the carbon intensity of electricity was viewed as constant, every unit of energy was the same, and a unit improvement in computing's energy efficiency meant a proportional improvement in computing's carbon efficiency. 
As the industry aggressively optimized for computing's energy efficiency---driven by the need to scale while reducing operational costs---it was only serendipitous that a cost-driven approach was also the environmentally conscious choice. 

\begin{figure}[t]
  \centering
    \includegraphics[width=\linewidth]{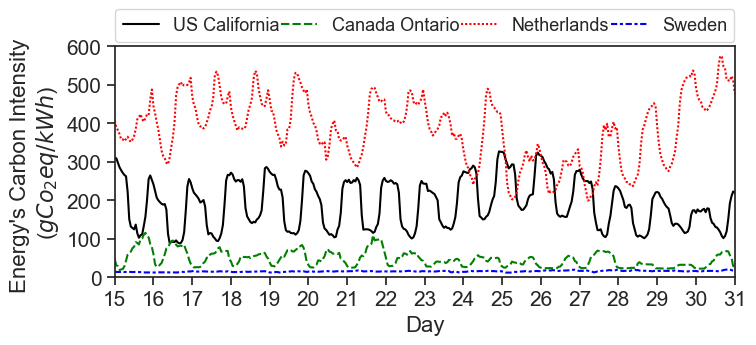}
    \vspace{-0.7cm}
    \caption{\emph{Energy's carbon intensity (05/15/2022 - 05/31/2022).}}
    \label{fig:trace_sample}
    \vspace{-0.5cm}
\end{figure}

However, the evolution of the electric grid over the last decade has diversified the mix of energy sources used for electricity generation. 
With a higher penetration of renewable energy in the electric grid and advancements in traditional power plant technologies, such as combined heat and power (CHP) plants, the carbon intensity of electricity now varies widely over time and across locations~\cite{iea-capacity, tammy-arxiv}.
Figure \ref{fig:trace_sample} illustrates the carbon intensity of energy in $gCO_2eq/kWh$ for four locations around the globe. 
Sweden exhibits a very low carbon intensity due to its reliance on hydropower, while the Netherlands has a {higher} carbon intensity due to its fossil fuel-heavy resource mix. 
Furthermore, Ontario and California experience diurnal changes in carbon intensity due to the increased use of solar energy. 
The growing penetration  of renewables in the electric grid has decreased the carbon intensity worldwide but also highlighted the importance of considering the timing and location of energy consumption. 
Computing workloads offer the flexibility to choose when and where to execute and consume energy. 
However, mechanisms that enable the exploitation of computing's flexibility tend to be energy-inefficient. 
Therefore, in most cases, achieving carbon efficiency requires sacrificing energy efficiency.

From a business standpoint, it does not make financial sense to be purposefully energy-inefficient as it costs money, especially in the absence of penalties on carbon emissions. 
However, there is a social incentive to reduce carbon footprint, and the computing industry is responding to this problem in two ways. 
First, assuming a time-varying carbon intensity, operators are leveraging the performance flexibility of workloads such as delaying or relocating workloads. 
Second, the industry is using various forms of carbon credits and offsets to reduce its estimated carbon emissions. 
Initially, carbon credits and power purchase agreements were used to offset carbon emissions on an annual basis. 
However, recently, the industry has started using more stricter forms of carbon offsets that match the energy demand of datacenters with renewable energy generation on the same distribution grid on an hourly basis, known as 24/7 matching~\cite{google-cfe}. 
While this is a step in the right direction, it should not be construed as running the datacenters purely on zero-carbon energy because datacenters still rely on the electric grid. 
As the electric grid still uses carbon-intensive energy sources, no one can be fully carbon-free until the grid itself is carbon-free. 
In this futuristic world, the focus can shift back toward the gains in energy efficiency that the industry has helped achieve.
In the meanwhile, blindly focusing on energy efficiency leaves many possible carbon-specific optimizations on the table. 
Unfortunately, just like the tension between alternating and direct current in the late nineteenth century, this debate has become a war of narratives and financial levers rather than a technical one~\cite{war-of-currents, war-of-currents-wiki}.
As grids worldwide still rely on carbon-emitting energy generation sources, we need to exploit all the flexibility of computing workloads such that we maximize carbon efficiency while optimizing energy efficiency. 

Computing workloads, depending on the application, offer flexibility along multiple dimensions. 
They can be delayed, paused, and resumed (\textit{temporal flexibility}). 
They can be assigned more or fewer resources (\textit{resource scaling}). 
They can be scaled up or down using Dynamic Voltage and Frequency Scaling (DVFS) (\textit{rate shifting}). 
Finally, they can be executed at a different geographical location (\textit{spatial shifting}). 
These mechanisms for exploiting flexibility could be energy-inefficient to varying degrees and yield different amounts of carbon savings. As energy inefficiency costs money, we need to analyze the trade-offs between carbon efficiency and energy efficiency.
This can not only guide the industry in estimating the cost of optimizing for carbon but also help regulators determine the appropriate incentives and penalties for carbon emissions. 

The management of carbon emissions in cloud datacenters is receiving significant attention due to the growing impact of climate change~\cite{ecovisor,zero-carbon,gupta2021chasing,embodied2,dean-carbon,arxiv-google,wait-awhile,cloudcarbon,tammy-arxiv}.  
While some studies have focused on embodied carbon, which refers to carbon emissions from the manufacturing and relocation of infrastructure, we are concentrating on the operational energy and carbon footprint of powering and cooling the infrastructure. Although our efficiency metrics can encapsulate other accounting methods, such as embodied carbon, by spreading the embodied carbon over the actual lifespan of the infrastructure, we did not use that combination since it does not comply with the GHG protocol~\cite{ghg} as highlighted by other researchers~\cite{hotcarbon-embodied}.

To the best of our knowledge, our work is the first to explicitly quantify the energy-carbon trade-offs of various flexibility mechanisms. To demonstrate this trade-off, we consider real-world carbon traces and analytically-modeled applications and simulate the effect of carbon-saving mechanisms. We use a state-of-the-art energy-efficient execution as the baseline and demonstrate how carbon efficiency can be significantly increased by being energy inefficient, and blindly optimizing for energy efficiency is not always the right approach. We also highlight the trade-off breadth of different techniques and show that there exists a tipping point where carbon savings is not yet affected by the energy inefficiency of such mechanisms. Beyond this point, the carbon footprint of energy overheads overweighs the reduction in carbon savings from exploiting flexibility.

\vspace{-0.2cm}
\section{Illustrating Efficiency Trade-offs}
\label{sec:mechanisms}
In this section, we investigate the trade-off between carbon efficiency and energy efficiency of four commonly-used mechanisms for exploiting computing's flexibility: temporal shifting, resource scaling, rate shifting, and spatial shifting. We choose state-of-the-art energy-efficient execution as the baseline unless stated otherwise. Therefore, our carbon efficiency gains come from optimizing specifically for carbon and not from improving computing's energy efficiency.  Furthermore, the term ``\emph{carbon efficiency}'' refers to computing's carbon efficiency and not energy's carbon efficiency, which is referred to by the reciprocal term of carbon intensity. 

We use a 3-year-long carbon intensity trace for Ontario, Canada, from electricity map~\cite{electricity-map} spanning January 1, 2020, to December 31, 2022. The trace provides the hourly average intensity values, measured in $gCO_2eq/kWh$. Unless stated otherwise, we assume that the job starts at the hour boundary, and aggregate our results across jobs starting at each hour of the year. We report carbon efficiency and energy efficiency values, with the results normalized to the most energy-efficient for energy efficiency and the least carbon-efficient for carbon efficiency, unless otherwise specified. Furthermore, we present our analysis across a wide range of empirically-driven configurations that map to real-world server classes and application characteristics to ensure our results are broadly applicable and not tied to a particular application or hardware. 

\vspace{-0.2cm}
\subsection{Temporal Shifting}
\label{sec:time-shifting}
The time-varying nature of electricity's carbon intensity creates green time periods, where the carbon intensity is significantly lower than the average carbon intensity for that location, as shown in Figure~\ref{fig:trace_sample}. 
The simplest, and most common, strategy to increase computing's carbon efficiency is to wait for such low carbon periods to arrive, execute the job during a given period, suspend the job at the end of this period, and resume its operation during the next low carbon period. 
However, this intermittent execution of jobs leverages checkpoint and restore techniques to save the state between low carbon periods, incurring energy overhead. 
The amount of overhead depends on the frequency of checkpoint and restore operations and the energy cost of a single checkpoint and restore operation, which, in turn, depends on the size of the state of the application~\cite{flint}. 

\begin{figure}[t]
  \centering
  \begin{subfigure}{0.8\linewidth}
        \centering
      \includegraphics[width=0.9\textwidth]{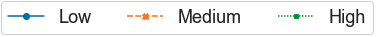}
  \end{subfigure}
  \\
  \begin{subfigure}[t]{0.48\linewidth}
    \includegraphics[width=\textwidth]{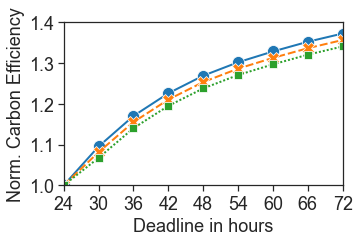}
    \vspace{-0.4cm}
    \caption{Carbon Efficiency}
    \label{fig:timeshift_carb_syn}
  \end{subfigure}
  \hfill
  \begin{subfigure}[t]{0.48\linewidth}
    \includegraphics[width=\textwidth]{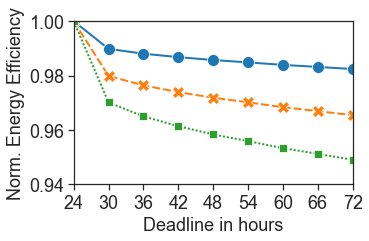}
    \vspace{-0.4cm}
    \caption{Energy Efficiency}
    \label{fig:timeshift_energy_syn}
  \end{subfigure}
    \vspace{-0.4cm}
  \caption{\emph{Carbon and energy efficiency of jobs with different overheads. Longer deadlines allow a higher use of checkpoint \& restore to operate in low carbon periods.}}
  % for synthetic profile
  \label{fig:timeshift_syn}
    \vspace{-0.2cm}
\end{figure}

Prior work on temporal shifting to reduce the carbon footprint has focused on configuring the thresholds for identifying low carbon periods, with or without future knowledge of carbon intensity, and the trade-offs between carbon savings and job completion times~\cite{ecovisor, radovanovic2021carbonaware, wait-awhile}, while ignoring the overhead of suspend-resume. 
There is also a recent work that explores online algorithms, taking into account the overhead of suspend and resume methods~\cite{lechowicz2023online}. 
However, these studies do not explicitly discuss or quantify the trade-off between the loss of energy efficiency and gains in carbon efficiency. 
We bridge this gap in an empirical study outlined next.

\noindent\emph{{\textbf{1. Experimental Setup.}}} Our setup presumes multiple jobs with different overhead percentages, aiming to optimize carbon efficiency using state-of-the-art carbon-aware execution policies. 

\noindent
\textbf{Applications.}
We consider an application that constantly performs computation, such as ML training, and requires a certain memory size to store the intermediate results.  
We assume that the job has performance flexibility and allows the operator to checkpoint \& restore its state. 
We use three variants of this job represented by their checkpoint \& restore overheads, which we configure as the time it takes to checkpoint or restore the memory state of the job. 
We set the overheads for the three variants as 5 minutes (low), 10 minutes (medium), and 15 (high) minutes, which can be mapped to real-world applications by assuming ML training over different-sized models.
The job also has temporal flexibility, aka slack, which we define as a multiple of the job's uninterrupted execution time. 
For example, a slack factor of 1.5$\times$ for a 24hrs job without interruption means that it has 36hrs to finish.
In our experiments, we consider a 24hrs long job and vary the slack factor from 1$\times$ to 3$\times$ of the job's runtime.

\noindent
\textbf{Policy.} 
We use a deadline-aware \emph{suspend-resume} policy to execute the job that has been proposed in recent work to reduce the carbon footprint of jobs with temporal flexibility~\cite{wait-awhile}. 
This policy assumes perfect future knowledge of carbon intensity and selects low carbon slots for executing the job such that it finishes before the deadline. 
It does not take into account the energy overhead of checkpoint \& restore when determining the number and selection of slots for execution.
However, we take into account the carbon overhead of intermittent execution and subtract that from carbon savings when calculating carbon efficiency gains.

\noindent\emph{{\textbf{2. Experimental Results.}}}
The results in Figure~\ref{fig:timeshift_carb_syn} show that a higher degree of flexibility (higher slack) can lead to greater reductions in carbon emissions (increased carbon efficiency). 
However, flexibility comes at the cost of energy efficiency. 
As shown in Figure~\ref{fig:timeshift_energy_syn}, larger slacks allow applications to checkpoint \& restore more often, increasing the energy and carbon overhead, which reduces energy efficiency. In Figure~\ref{fig:timeshift_energy_syn}, the y-axis limit is set between 1 and 0.94 to ensure clear visibility of the lines representing normalized energy efficiency.
It is important to note that the magnitude of carbon savings and the impact on energy efficiency are application-specific, but their relationship is fundamental and will hold across application characteristics and carbon intensity profiles.

\noindent
\emph{\textbf{3. Key Takeaways.}
Higher temporal flexibility enables applications to increase their carbon efficiency, but the gains depend on how often applications incur the energy and carbon overhead of the checkpoint and restore mechanism to take advantage of low carbon periods.
}

\vspace{-0.2cm}
\subsection{Resource Scaling}
\vspace{-0.05cm}
\label{sec:resource_scaling}
Resource scaling is the method of adding or removing resources to a given job to speed up or slow down the speed of execution, respectively. 
In the context of carbon-aware computing, resource scaling can be used as an antidote to the increase in job completion time under \emph{suspend-resume} execution~\cite{ecovisor, hanafy2023carbonscaler}. 
Instead of resuming the job at 1$\times$ during low carbon periods, it can be scaled up to k$\times$ to make up for the time spent in suspend state. 
However, the effectiveness of scaling depends on the application characteristics, such as the size of sequential barriers modeled by Amdahl's law~\cite{amdahls-law}.   
As a result, as the allocated resources increase, the speed up increases sub-linearly, reducing the energy efficiency of execution.

\begin{figure}[t]
  \centering
  \begin{subfigure}{0.8\linewidth}
        \raggedright
      \includegraphics[width=\textwidth]{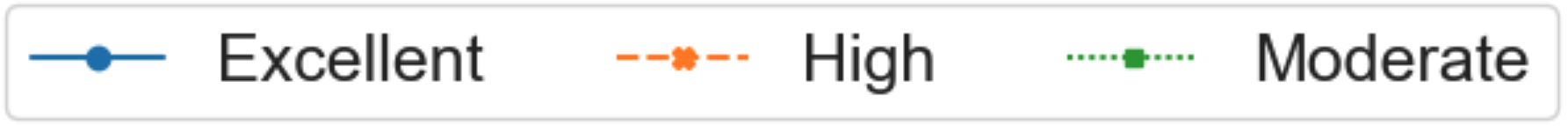}
  \end{subfigure}
  \\
  \begin{subfigure}[t]{0.48\linewidth}
    \includegraphics[width=\textwidth]{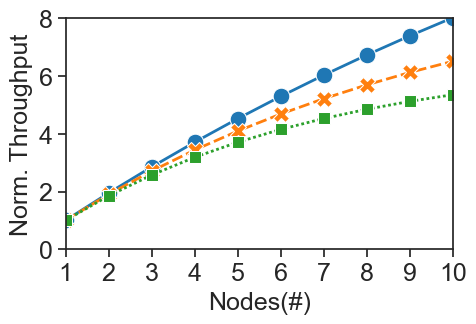}
    \vspace{-0.5cm}
    \caption{Throughput}
    \label{fig:scale_throughput}
  \end{subfigure}
  \hfill
  \begin{subfigure}[t]{0.48\linewidth}
    \includegraphics[width=\textwidth]{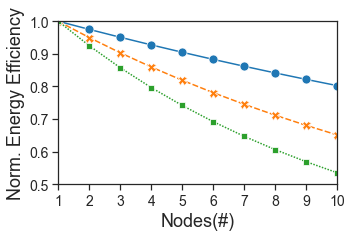}
    \vspace{-0.5cm}
    \caption{Energy Efficiency}
    \label{fig:scale_efficiency}
  \end{subfigure}
   \vspace{-0.3cm}
  \caption{\emph{As applications with sub-linear scaling characteristics scale, their energy efficiency reduces.}}
  \label{fig:scaling_applications}
  \vspace{-0.6cm}
\end{figure}

\noindent\emph{{\textbf{1. Experimental Setup.}}}
In this experiment, we demonstrate the carbon efficiency and energy efficiency trade-off for an application that leverages resource scaling to reduce its carbon footprint. 

\noindent\textbf{Applications.} Since applications' scalability dictates the energy overhead, we consider different scalability in terms of the throughput reduction per additional node. 
Figure \ref{fig:scaling_applications} shows the scalability characteristics for three applications with excellent scaling (5\% reduction in normalized throughput per additional node), high scaling (10\% reduction per additional node), and moderate scaling (15\% reduction per additional node) characteristics. This behavior is similar to many real-world distributed applications as shown in \cite{qi17paleo}.
We assume that the energy consumption of each additional node is the same. 
As a result, as shown in Figure \ref{fig:scale_efficiency}, the energy efficiency of the computing decreases due to reduced normalized throughput after scaling.

\noindent
\textbf{Policy.}
\citet{ecovisor} propose a carbon-aware scaling policy called \emph{Wait\&Scale}, which selects an application-specific scale factor based on its scalability characteristics. 
It assumes perfect knowledge of future carbon intensity.
Similar to a carbon-aware suspend-resume policy, it generates a schedule that suspends the job during high carbon periods and resumes it at the application-specific scale factor during the low carbon periods.
In this scenario, the scale factor is set to 1, which means that the job completion time is the same as the completion time for an uninterrupted execution. 
As a result, jobs do not gain carbon savings from temporal flexibility. 

\begin{figure}[t]
  \centering
  \begin{subfigure}{0.48\textwidth}
        \raggedright
       \includegraphics[width=\textwidth]{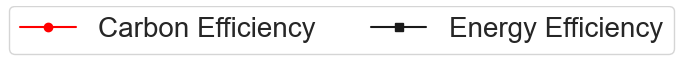}
    \end{subfigure}
    \\
  \begin{subfigure}[b]{0.45\textwidth}
    \includegraphics[width=\textwidth]{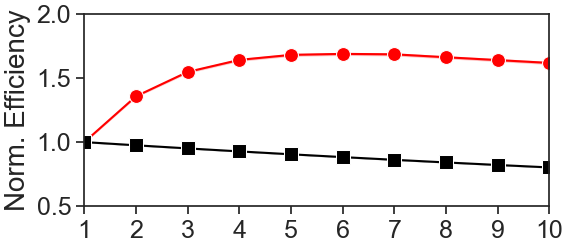}
    \vspace{-0.5cm}
    \caption{Excellent}
    \vspace{-0.1cm}
    \label{fig:scale-saving-excellent}
  \end{subfigure}
  \\
  % \quad % some horizontal space between the subfigures
  \begin{subfigure}[b]{0.45\textwidth}
    \includegraphics[width=\textwidth]{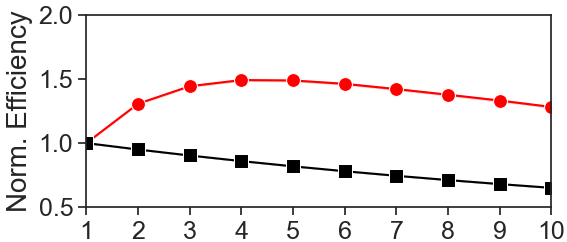}
    \vspace{-0.5cm}
    \caption{High}
    \vspace{-0.1cm}
    \label{fig:scale-saving-high}
  \end{subfigure}
  \\
  % \quad % some horizontal space between the subfigures
  \begin{subfigure}[b]{0.45\textwidth}
    \includegraphics[width=\textwidth]{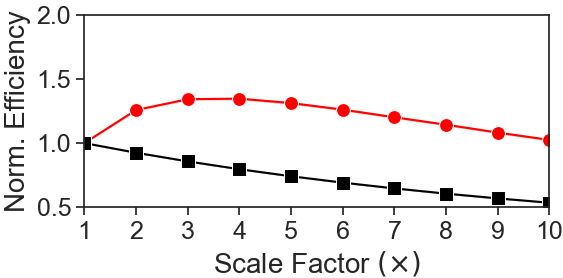}
    \vspace{-0.4cm}
    \caption{Moderate}
    \label{fig:scale-saving-moderate}
  \end{subfigure}
  \vspace{-0.4cm}
  \caption{\emph{Normalized carbon and energy efficiency for different scalability profiles when running in Ontario, Canada.}}
  \label{fig:scale-saving}
  \vspace{-0.6cm}
\end{figure}

\noindent
\textbf{\emph{2. Experimental Results.}}
Figure \ref{fig:scale-saving} shows the normalized carbon efficiency and energy efficiency of the three scalable applications with scalability characteristics shown in Figure \ref{fig:scaling_applications}. 
The results indicate that maximizing energy efficiency does not necessarily result in maximum carbon efficiency. 
Rather, the increase in carbon efficiency depends on the flexibility that comes with a decrease in energy efficiency. 
Furthermore, the results in Figure \ref{fig:scale-saving} demonstrate that the application's scalability plays a significant role in carbon efficiency gains and energy efficiency losses.
For instance, the job with excellent scaling (Figure \ref{fig:scale-saving-excellent}) was able to increase its carbon efficiency by 68\% at an energy efficiency loss of 15\%. Conversely, the moderately scalable job showed only a 34\% increase in carbon efficiency but paid more than a 25\% loss in energy efficiency. 

It is also worth noting that, in all three cases, Figure \ref{fig:scale-saving} illustrates that increasing the scaling factor does not always improve carbon efficiency. 
Beyond a certain scaling factor, the gain in throughput during high carbon periods does not overcome the carbon cost due to low energy efficiency at high scales. 
Hence, the reciprocal behavior that loss of energy efficiency does not always lead to carbon efficiency should also be considered while scaling applications. 

\noindent
\emph{\textbf{3. Key Takeaways.} Carbon-aware application of scaling policies can yield significant gains in carbon efficiency. However, the high energy overhead of scaling means that applications must be highly judicious in choosing their scale factor as gains become marginal at high scales.}

\subsection{Rate Shifting}
\label{sec:rate-shifting}
Dynamic Voltage and Frequency Scaling (DVFS) has been widely used for energy optimization for servers in datacenters by leveraging the non-linear relationship between power consumption and application throughput~\cite{datacenter-dvfs, weiser-dvfs, heracles, flex}.
DVFS can also be used for optimizing carbon efficiency where the application runs faster using a higher  frequency during low carbon periods and saves energy by lowering the CPU frequency and its execution speed during high carbon periods, possibly at a lower energy-efficiency. 
%During high carbon periods, instead of completely suspending its operation, the application can operate at a frequency that makes slow progress at higher energy efficiency. 
DVFS can especially be helpful for uninterruptible applications that cannot be suspended, as it makes progress at all times without suspension. 

The energy savings in using DVFS come from the non-linear relationship between a processor's power demand ($P$) and its frequency ($f$) and voltage ($V$) governed by the following equation,  
\begin{equation}\label{eq:power-demand}
    P = CfV^2+P_{\texttt{static}}.
\end{equation}

Here, $C$ and $P_{\texttt{static}}$ are processor architecture-specific constants. 
As shown in this equation, power demand has a dynamic and a static range. The dynamic range is dictated by the linear relation with its frequency; reducing the operating frequency by 50\% will reduce the dynamic range by 50\%. However, higher savings in the dynamic range come from the non-linear relationship with the operating voltage of the processor; decreasing the operating voltage to half reduces the dynamic power consumption of the processor by a factor of 4.
It is worth mentioning that a decrease in the power consumption of a processor does not lead to a proportional decrease in its performance. 
The decrease in performance, denoted by $S$, depends on the CPU-boundedness of applications and can be modeled using Amdahl's law~\cite{amdahls-law} by the following equation:
\begin{equation}\label{eq:scale-down}
    S = \frac{1}{io+ \sfrac{(1-io)}{f_n'}}.
\end{equation}
where $io$ is the fraction of time an application spends accessing Input/Output (IO) peripherals at the maximum operating frequency ($F_{max}$), which is one of the major reasons affecting the application's slowdown due to their sequential nature. 
$f_n'$ is the normalized frequency w.r.t. the highest frequency $f_n' = f_n/F_{max}$, where $f_n\in [F_{\min}, F_{\max}]$. 

Considering only the dynamic range, by assuming $P_{static}$ to be 0 and $C$ to be 1, a 50\% decrease in operating frequency for an application with 50\% IO yields only a 34\% reduction in the application's performance, making it more efficient. 
While the reduction of frequency alone yields 50\% savings in power, more savings will be achieved if the operating voltage is decreased as well. 
However, many modern processors do not allow direct and separate control of the processor's operating voltage. 
Instead, changing the operating frequency alters the operating voltage to a pre-determined voltage level. 
As a result, applications may not have a full range of parameters available to them to optimize energy efficiency using DVFS. 
Furthermore, depending on the practical values for $P_{static}$ and $C$, energy efficiency gains may be further limited. 

\noindent
\textbf{\emph{1. Experimental Setup.}}
We next evaluate the impact of DVFS for different application characteristics under different operation frequencies to highlight the tension between carbon and energy efficiency.

\begin{figure}[t]
\centering
\includegraphics[width=\linewidth]{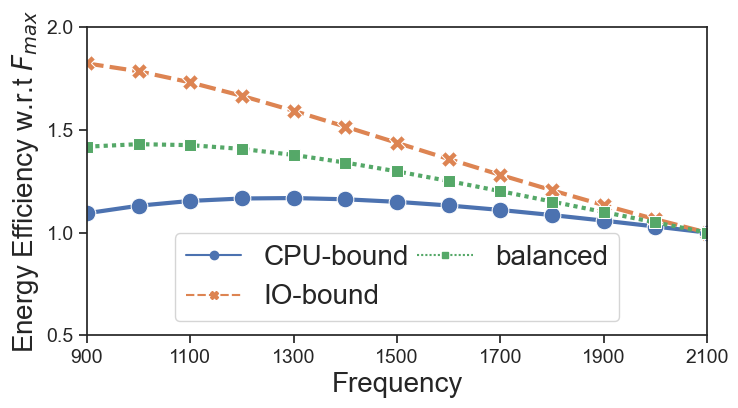}
\vspace{-0.7cm}
\caption{\emph{Normalized energy efficiency across different application profiles defined by their IO-intensiveness.}}%
\label{fig:dvfs_energy_efficiency}
\vspace{-0.5cm}
\end{figure}

\noindent
\textbf{Server Configurations.}
Figure \ref{fig:dvfs_energy_efficiency} shows our modeled server, modeled after a local server in our lab, that has an Intel-Xeon Processor E5-2620 v4 with DVFS enabled. The server dynamic power range was controlled through the frequency range of (0.9 to 2.1GHz, 0.1 MHz step), voltage range of 0.8 to 1.2V. The CPU's transistor gate's capacitance ($3\times10^{-2}$), while the static power of the server is 30W, which is approximately 25\% of its operational power.
In practice, frequency levels and voltage values are tied together. Thus, we split the voltage range into the same number of steps as frequency and establish a direct relationship between frequency and voltage levels.

\noindent
\textbf{Applications.}
We model three different applications that map to real-world: ``CPU-bound'' application such as matrix multiplication in ML training (0\% time spent on input/output), ``IO-bound''application of text processing, e.g., Hadoop (70\% time spent on I/O), and ``balanced'' application such as in-memory data processing, e.g., Spark, (40\% time spent on I/O). 
In all of these examples, I/O time is configured when the processor is running at the highest frequency. 
The application-specific energy efficiency profile is created based on its I/O\%, under different frequencies, by equations \ref{eq:power-demand} and \ref{eq:scale-down}.

\noindent
\textbf{Policy.}
There is no prior work on leveraging DVFS to optimize for carbon efficiency. 
Therefore, we devise a simple strategy to explore the carbon and energy efficiency trade-off. 
Our policy uses the application-specific profile to operate at a high frequency, often less  energy efficient, during low carbon periods and operate at a low frequency, e.g.,  highest energy efficient frequency, during high carbon periods. 
The complex decision space necessitated evaluating the policy against all frequency permutations. In this case, the policy is given all permutations of $F_1, F_2 \in [900,...,2100]$, and we also use the mean carbon intensity $\mu_c$, during the expected execution period, as the threshold. In this case, at a time slot $i$, the application runs at frequency $F_1$ if the carbon intensity $c_i$ is less than or equal to the threshold $\mu$ ($c_i \le \mu$), and it runs at frequency $F_2$ otherwise.

\noindent
\textbf{\emph{2. Experimental Results.}}
We model the energy efficiency behavior of our applications by computing the normalized energy consumption at the highest frequency with respect to other frequencies. Figure \ref{fig:dvfs_energy_efficiency} shows that across application classes, I/O-bound applications observe the most gains since an I/O-heavy application does not utilize the CPU to its full extent and is not affected by the slower processing speed of the CPU at lower frequencies, while a CPU-bound application barely sees any gains as its throughput is highly affected by the processing speed. The efficiency gains from DVFS were further in prior work~\cite{dvfs_diminishing}.

\begin{figure}[t]
\centering
\begin{subfigure}{0.49\textwidth}
  \centering
  \begin{subfigure}[b]{0.48\textwidth}
    \includegraphics[width=\textwidth]{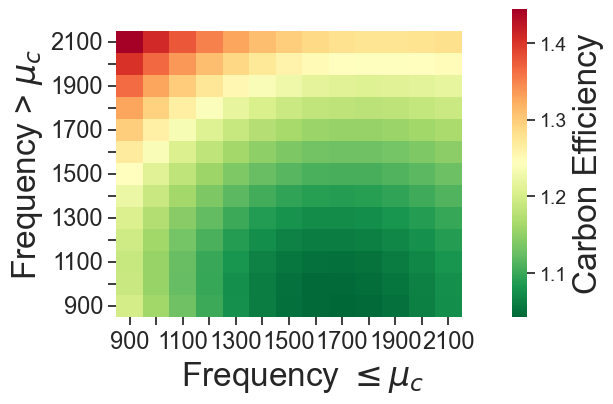}
    %\caption{Carbon}
    \label{fig:dvfs-results-Server2-cpu-bound-carbon}
  \end{subfigure}%
  \begin{subfigure}[b]{0.48\textwidth}
    \includegraphics[width=\textwidth]{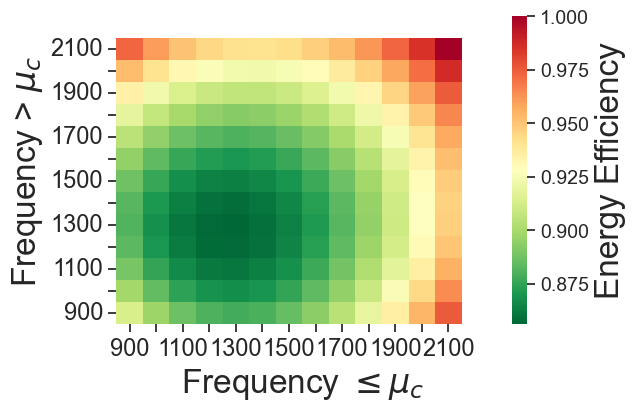}
    %\caption{Energy}
    \label{fig:dvfs-results-Server2-cpu-bound-energy_efficiency}
  \end{subfigure}%
  \vspace{-0.4cm}
  \caption{CPU-bound Task}
  \label{fig:dvfs-results-Server2-cpu-bound}
\end{subfigure}%
\\
\begin{subfigure}{0.49\textwidth}
  \centering
  \begin{subfigure}[b]{0.48\textwidth}
    \includegraphics[width=\textwidth]{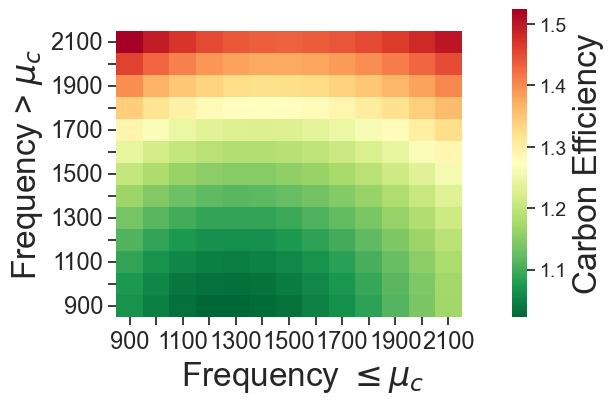}
    %\caption{Carbon}
    \label{fig:dvfs-results-Server2-balanced-carbon}
  \end{subfigure}%
  \begin{subfigure}[b]{0.48\textwidth}
    \includegraphics[width=\textwidth]{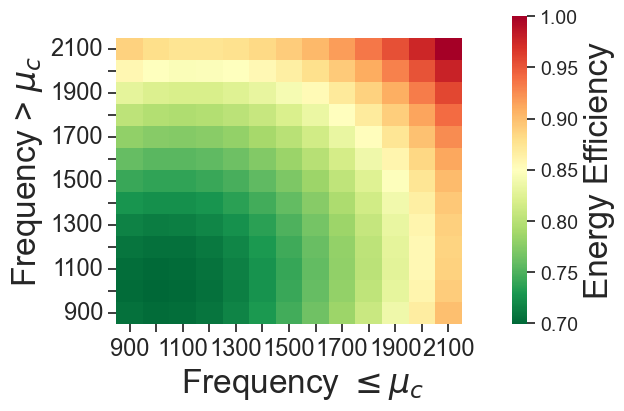}
    %\caption{Energy}
    \label{fig:dvfs-results-Server2-balanced-energy}
  \end{subfigure}%
  \vspace{-0.4cm}
  \caption{Balanced Task}
  \label{fig:dvfs-results-Server2-balanced}
\end{subfigure}%
\\
\begin{subfigure}{0.49\textwidth}
    \centering
  \begin{subfigure}[b]{0.48\textwidth}
    \includegraphics[width=\textwidth]{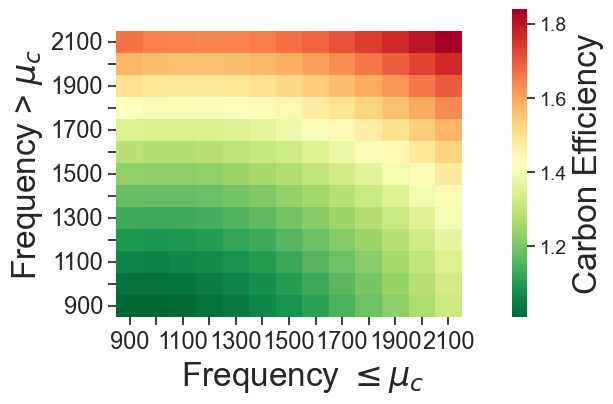}
    %\caption{Carbon}
    \label{fig:dvfs-results-Server2-io-bound-carbon}
  \end{subfigure}%
  \begin{subfigure}[b]{0.48\textwidth}
    \includegraphics[width=\textwidth]{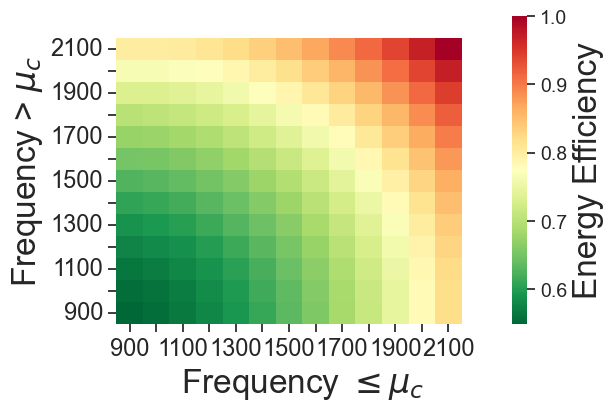}
    %\caption{Energy}
    \label{fig:dvfs-results-Server2-io-bound-energy}
  \end{subfigure}%
  \vspace{-0.4cm}
  \caption{IO-bound Task}
  \label{fig:dvfs-results-Server2-io-bound}
\end{subfigure}
\vspace{-0.6cm}
\caption{\emph{Carbon and energy efficiency of different tasks.}}
\vspace{-0.6cm}
\label{fig:dvfs-results-Server2}
\end{figure}

Figure~\ref{fig:dvfs-results-Server2} shows the carbon and energy efficiency, of the three applications denoted as CPU-bound (0\% IO), IO-Bound (70\% IO), and balanced (40\% IO). The results clearly indicate that energy-efficient configurations do not always yield the highest carbon efficiencies. For example, Figure \ref{fig:dvfs-results-Server2-cpu-bound} shows that the highest carbon efficiency is achieved by running fast ($\sim$1.7GHz) when energy's carbon intensity is low while running slow ($\sim$1GHz) otherwise, contrarily, always running at ($\sim$1.3GHz) yields highest energy efficiency. Figure \ref{fig:dvfs-results-Server2-io-bound} shows another example where energy and carbon efficiency are more correlated as the energy and carbon efficiency increase from lowering the frequency. We point out that other configurations resulted in similar conclusions, but we had to leave them out due to space constraints.

\noindent
\emph{\textbf{3. Key Takeaways.} Using DVFS in a carbon-aware manner can greatly improve carbon efficiency. Adjusting the operating frequency can result in high-efficiency gains that might bridge the gap between energy and carbon-efficient computing.}

% \vspace{-1ex}
\subsection{Spatial Shifting}
\label{sec:spatial-shifting}
Migrating services across the network in order to optimize cost or latency has been widely discussed~\cite{wan-migration, follow-the-sun, zero-carbon-network, tammy-arxiv}. Similarly, migrations can be utilized to increase carbon efficiency by transferring jobs to a region where the carbon intensity of energy is lower. For example, a task can be migrated across the globe by following the availability of carbon-free solar energy. 
However, flexibly transferring jobs between locations decreases energy efficiency as it involves energy overheads from checkpointing, transferring, and restoring execution state as well as application data. For instance, for migrating a data processing application (e.g., ML training), the application state must be checkpointed at the source, moved across the network, and restarted at the destination. Also, the data must be moved or cloned in both locations which incurs extra energy consumption.  The process of checkpointing and restoring depends on the application size as explained in section \ref{sec:time-shifting}, while the migration depends on the state and data size, and cloning also comes with energy and financial cost overheads.  A full analysis of spatial shifting 
trade-offs is left to future work.

\vspace{-1ex}
\section{Conclusion}
\label{sec:discussion}
For a long-time, energy efficiency has been a key objective for cost-effective and sustainable computing. The necessity to decrease computing's operating costs and the environmental impact of computing made energy efficiency a first-class citizen in computing.
However, the wide adoption of clean energy in electrical grids, along with increasing public awareness about energy sources, and the enabled visibility of the time-varying carbon intensity, has resulted in a shift where the most energy-efficient operations may no longer be considered the most sustainable or socially acceptable choice. For these reasons, carbon efficiency (the amount of work per unit of carbon) appeared as a ``true'' sustainability metric. The key idea in increasing carbon efficiency is to exploit computing's workloads flexibility by adjusting execution time (Temporal Shifting), speed (Scaling and Rate Shifting), and location (Spatial Shifting) according to the grid's carbon intensity.  In this paper, we highlighted an inevitable tension between carbon and energy efficiency. We explored the core mechanisms used in carbon-efficient computing along with policies from the state-of-the-art in a wide range of scenarios. 
The paper demonstrated qualitatively and quantitatively that ``striving for maximum energy efficiency is not always the most sustainable (carbon-efficient) approach''. The gains and overheads of combining multiple carbon-aware flexibility mechanisms is left for future work.

\section*{Acknowledgements}
We thank the HotCarbon reviewers for their valuable comments, which improved the quality of this paper. We also thank WattTime and electricityMap for providing the carbon-intensity data. This research is supported by NSF grants 2213636, 2136199, 2106299, 2102963, 2105494, 2021693, 2020888, 2045641, as well as VMware.

\bibliographystyle{ACM-Reference-Format}

\end{document}